\documentclass{emulateapj}
\usepackage{footmisc}

\usepackage[bookmarks=false]{hyperref}
\usepackage{breakurl}
\usepackage{natbib}
\usepackage{amsmath}

\def\beq{\begin{equation}} 
\def\eeq{\end{equation}} 
\def\beqar{\begin{eqnarray}} 
\def\eeqar{\end{eqnarray}}

\def\pfrac#1#2{\left( \frac{#1}{#2} \right)} 
\def\avg#1{\langle #1 \rangle}

\def \nn{\nonumber}

\def\beq{\begin{equation}} 
\def\eeq{\end{equation}} 
\def\beqar{\begin{eqnarray}} 
\def\eeqar{\end{eqnarray}} 
\def\bal{\begin{align}}
\def\eal{\end{align}}

\def\pfrac#1#2{\left( \frac{#1}{#2} \right)} 
\def\avg#1{\langle #1 \rangle}

\def \nn{\nonumber}

\def \l({\left(}
\def \r){\right)}
\def \eps{\epsilon}

\def\avg#1{\langle #1 \rangle}

\def\pfrac#1#2{\left( \frac{#1}{#2} \right)}

\def\fermi{{\it  Fermi}\ }

\def\egamma{E_{\rm \gamma}}

\def\sc#1{{\mathcal #1}}

\def\ao{\al_{1}}
\def\al{\alpha}

\def\dd#1#2{\frac{d #1}{d #2}}

\def\G#1#2{\Gamma_{\rm #1}(#2)}
\def\me{m_{\rm e}}

\def\nsig{n_{\rm \sigma}}
\def\rct{{\ \rm counts\ sec^{-1}}}
\def\reflux{{\ \rm ergs\ cm^{-2}\ sec^{-1}}}

\def\fq{F_{\rm q}}
\def\ff{F_{\rm f}}
\def\fm{\avg{F}}

\def\deleobsi#1{\Delta E_{\rm obs}(#1)}
\def\delepoli#1{\Delta E_{\rm pol}(#1)}

\def\eobsmini#1{E_{\rm obs,min}(#1)}
\def\eobsmaxi#1{E_{\rm obs,max}(#1)}
\def\epolmini#1{E_{\rm pol,min}(#1)}
\def\epolmaxi#1{E_{\rm pol,max}(#1)}

\def\fphpolcal{F_{\rm ph,\ \Delta E_{\rm pol},calib} }

\def\feobscali#1{F_{\rm E,\ \Delta E_{\rm obs}, calib}(#1)}
\def\fepolcali#1{F_{\rm E,\ \Delta E_{\rm pol},calib}(#1)}

\def\fphpolcali#1{F_{\rm ph,\ \Delta E_{\rm pol},calib}(#1)}

\def\fphpol{F_{\rm ph,\ \Delta E_{\rm pol}} }

\def\fphpolbl{F_{\rm ph,\ \Delta E_{\rm pol}, blazar}}

\def\feobsbli#1{F_{\rm E,\ \Delta E_{\rm obs}, blazar}(#1)}
\def\fepolbli#1{F_{\rm E,\ \Delta E_{\rm pol},blazar}(#1)}

\def\fphpolbli#1{F_{\rm ph,\ \Delta E_{\rm pol}, blazar}(#1)}

\def\rcrab{R_{\rm calib} }
\def\rbl{R_{\rm blazar}}
\def\mdpcrab{MDP_{\rm calib}}

\def\tobscrab{T_{\rm obs,calib}}
\def\tobsbl{T_{\rm obs,blazar}}
\def\tobs{T_{\rm obs}}
\def\tobsblbgd{T_{\rm obs,blazar}\bigg|_{\rm bgd\ dom}}
\def\tobsblcalsig{T_{\rm obs,calib}\bigg|_{\rm sig\ dom}}

\def\etaf{\eta_{\rm flare}}
\def\etaeff{\eta_{\rm flare,eff}}

\def\b#1{\beta_{\rm #1}}
\def\fb{F_{\rm b}}
\def\fs{F_{\rm sens}}
\def\cq{C_{\rm q}}
\def\nun{N_{\rm u}}

\def\ndet{N_{\rm det}}
\def\nqdet{N_{\rm q,det}}
\def\nfdet{N_{\rm f,det}}

\def\dndfo{ \dd{N}{F}\bigg|_{\rm obs} }

\def\etaavg{\eta_{\rm avg}}

\begin{document}

\title{High energy polarization of blazars: Detection prospects}

\author{N. Chakraborty\altaffilmark{1,2}, V. Pavlidou\altaffilmark{3,4}, B.D. Fields\altaffilmark{2}}

\altaffiltext{1}{Max-Planck-Institut f\"ur Kernphysik, Saupfercheckweg 1, 69117 Heidelberg, Germany}
\altaffiltext{2}{Department of Astronomy and Department of Physics, University of Illinois, Urbana, IL}
\altaffiltext{3}{Department of Physics, University of Crete, 71003 Heraklion,Greece}
\altaffiltext{4}{Foundation for Research and Technology - Hellas, IESL, Voutes, 7110 Heraklion, Greece}

\begin{abstract}

Emission from blazar jets in the ultraviolet, optical, and infrared is polarized. If these low energy photons were inverse-Compton scattered, the up-scattered high energy photons retain a fraction of the polarization. Current and future X-ray and gamma-ray polarimeters such as INTEGRAL-SPI, PoGOLITE, X-Calibur, GAP, GEMS-like missions, ASTRO-H, and POLARIX  have the potential to discover polarized X-rays and gamma-rays from blazar jets for the first time. 
Detection of such polarization will open a qualitatively new window into high-energy blazar emission;
actual measurements of polarization degree and angle will quantitatively
test theories of jet emission mechanisms. 
We examine detection prospects of blazars by these polarimetry missions using 
examples of 3C 279, PKS 1510-089 and 3C 454.3, bright sources with relatively high degrees of 
low energy polarization. We conclude that while balloon polarimeters will be challenged to detect blazars within reasonable observational times (with X-Calibur offering the most promising prospects), space-based missions should detect the brightest blazars for polarization fractions down to few percent. 
Typical flaring activity of blazars could boost the overall number of polarimetric 
detections by nearly a factor of 5-6 purely accounting for flux increase 
of the brightest of the
comprehensive, all-sky, \fermi-LAT blazar distribution.  
Instantaneous increase in the number of detections is approximately a factor of 
2 assuming duty cycle of $20\%$ for every source. 
Detectability of particular
blazars may be reduced if variations in flux and polarization
fraction are anticorrelated.  Simultaneous use of variability and polarization trends
could guide the selection of blazars for high energy polarimetric observations. 

\end{abstract}

\section{Introduction}
\label{sect:intro}

Active Galactic Nuclei (AGNs)
are some of the most 
luminous yet mysterious
objects in the universe. 
Their particle and radiative emissions 
are powered by central supermassive black holes accreting matter. 
They are frequently observed to host relativistic jets, where  
bulk kinetic energy
is converted to nonthermal random kinetic energy of electrons,
radiation across the EM spectrum, and, possibly, particle emission
(ions and neutrinos). 
Thus, these particles and radiation are messengers 
of the extreme astrophysical conditions in the core of active 
galaxies and their jets. 

Blazars are AGNs where 
the observer's line of sight is closely aligned with the jet axis.
Various properties 
of the radiation from blazars like the overall intensity, spectrum,  and 
variability 
have been studied with multiwavelength observations.
They have a non-thermal spectral energy distribution,
with a low energy broadband peak  in the range of radio to UV or even
X-rays, and a second, high-energy peak, which starts from the X-ray
band and can reach TeV or even higher energies. 
Their low energy peak 
is well explained by synchrotron from relativistic leptons, and as
such can be highly polarized 
\citep[e.g., ][]{1986rpa..book.....R}.  If higher energy emission is in part or in whole due to inverse Compton interactions of the polarized synchrotron component, then it too will retain some polarization. 

While blazar polarization at low energies is well-established,
blazar polarization at high energies has received much less attention, 
due to a lack of data.
Indeed, there are few astrophysical sources of any kind
that have measured high-energy polarization.
Detections of polarized gamma-rays near the Crab Nebula and pulsar
\citep[e.g.,][]{2008Sci...321.1183D, 2013arXiv1302.3622M} 
are the most significant of the high energy polarimetric observations, 
with the Crab being used for calibrating polarization observations. 
Solar flares also have been observed in polarized 
X-rays \citep{2003SPD....34.2205M}. 
Amongst transient sources, there are very few successful observations of gamma-ray bursts (GRBs) 
as listed in \citet{2013ApJ...769...70C}. No high energy 
polarimetric data exists for blazars. This is in part due 
to the challenges in measurement of polarization in 
X-rays and soft gamma-rays. 
However, with numerous X-ray and soft gamma-ray polarimeters at various 
stages of planning, design and operation \citep[e.g.,][]{2010xrp..book.....B, 2012arXiv1211.5094P, 2013APh....41...63G, 2013arXiv1303.7158K, 2010ExA....28..137C, 2011AIPC.1358..408Y, 2012cosp...39.1259M} and 
studies of optical / FIR polarization properties of blazars underway 
\citep[e.g.,][]{2011PASJ...63..489S, 2011PASJ...63..639I},   
a systematic study of high energy polarization from blazars is 
attracting renewed interest \citep{Krawczynski:2011ym, 2013ApJ...774...18Z, 2013arXiv1303.7158K}. 

In this paper, we focus on detection prospects of 
X-ray and soft gamma-ray polarization of blazars with 
polarized seed photons in their jet. In Sec.~\ref{qualitative} we discuss qualitatively different models for the high-energy emission from blazars and the conditions under which this emission is expected to be polarized. 
 In Sec.~\ref{sect:degree}, 
we discuss quantitatively the degree of polarization expected from 
inverse-Compton scattering of polarized low energy photons 
by relativistic electrons in the jet with a power-law distribution. We 
estimate polarization values for three of the brightest blazars with 
fairly high degrees of polarization in the infra-red as an illustration. 
The minimum detectable polarization (MDP) for various 
telescopes in general is reviewed in Sec.~\ref{sect:sensitivity}. Sec.~\ref{sect:blazarlist} 
lists the chosen brightest blazars and the observing time required to 
detect the polarization for different polarimeters from the MDP formula. 
In section ~\ref{sect:flaring}, we discuss the 
influence of flaring on detection prospects guided 
by observational trends of polarization and flaring. 
We discuss our conclusions along with future work 
in section~\ref{sect:conc}. 

\section{High-energy Polarization in Blazars: Qualitative Discussion}\label{qualitative}

Blazar broadband spectral energy distributions exhibit two broad peaks. The first, lower energy peak
is usually attributed to synchrotron emission from 
leptons that provides a good explanation of the spectrum. 
This suggests that peak emission is intrinsically polarized \citep{1986rpa..book.....R}.
It ranges from radio to optical and UV, and, for some high-synchrotron-peaked blazars, to as high an energy as  
X-rays. Observations of flat spectrum radio quasars (FSRQs) 
and low energy peaked BL LAC (LBL) objects 
at low energies (radio to optical) do indeed confirm that the low-energy peak emission is
polarized\citep[e.g.,][]{2006NJPh....8..131L, 2004MNRAS.354...86R,
  1974ApJ...194..249W, 2010ApJS..189....1A, Fujiwara:2012gt, robopol}.The high energy peaked BL LACs on the other hand have 
their first peaks in the  
X-ray regime. X-ray polarimetry can help confirm 
the synchrotron origin of this peak. In this paper, 
we restrict ourselves to the blazars where 
polarimetric observations exist at low energies, i.e. to 
the FSRQs and LBLs. 

The second high-energy peak in blazar spectra 
is less well understood and a matter of active debate. It
is typically associated to inverse-Compton scattering of 
low energy photons. The emission can be purely leptonic, in which case primary accelerated electrons are the ones responsible for upscattering the lower-energy photons. Or it can involve hadronic processes as well. In the latter case, the primary accelerated particles also include protons and ions, and part of the high-energy emission is produced as a direct product of hadronic interactions (decay of neutral pions into gamma rays and proton synchrotron) and part through inverse Compton interactions of relativistic particles produced in hadronically-induced electromagnetic cascades. For the former case of inverse-Compton emission by primary electrons, seed photons can either originate in
external sources (External Compton mechanism, EC) 
  [e.g. \citet{2012ApJ...752L...4M} and earlier references therein] such as the
  accretion disk and the broad-line region or from internal sources within the
  jet like the synchrotron photons 
    [e.g. \citet{2012MNRAS.420...84Z} and earlier references therein] (self-synchrotron Compton, SSC). 

In the SSC case, the high-energy emission retains a fraction of the synchrotron photon polarization due to the nature of 
the inverse Compton scattering process \citep{Bonometto1970, 2009MNRAS.395.1507M, Krawczynski:2011ym, 2013ApJ...774...18Z}. The EC mechanism, on the other hand, is generally expected to produce lower degree 
of polarization from both analytical calculations \citep{Bonometto1970} and simulations \citep{Krawczynski:2011ym}. 
Simulations in \citet{2009MNRAS.395.1507M} find higher degrees of polarization for both EC and SSC. 

Polarization 
is a key ingredient in the multimessenger understanding of blazars. 
It can distinguish between different emission mechanisms 
of blazars such as hadronic and leptonic and 
even between different leptonic models 
as mentioned in, e.g., \cite{2009MNRAS.395.1507M, Krawczynski:2011ym, 2013arXiv1303.7158K} and demonstrated  
in a model-dependent way in \citet{2013ApJ...774...18Z}. 
Blazar variability can serve as an additional diagnostic for monitoring 
the relation between the jet activity and polarization. This 
will also allow measurement of polarization during the 
flaring states of those blazars 
whose polarization would, 
under normal circumstances, be undetectable. We examine 
the effect of flux increase due to flaring of blazars in general. The detailed study 
of relations between 
variability and polarization
 for individual blazars is left for future work.

\section{Degree of high energy polarization of blazars:  Expectations}
\label{sect:degree}

\subsection{High energy polarization from inverse-Compton scattering of low energy photons}
Polarization is quantified in terms of the degree of 
polarization, $\Pi$ which is the fraction of polarized light, 
and by the polarization position angle, $\psi$. 
\citet{Bonometto1970} calculated the polarization 
of photons produced by inverse-Compton scattering of 
 a monochromatic beam of photons 
by an unpolarized
electron distribution; both populations are assumed to have no
spatio-temporal variations. 
In this calculation, the initial and scattered photon energies ($\eps$
and $\egamma$ respectively) satisfy
$\egamma \gg \eps$
and 
$\egamma \eps/\me^2 \ll 1$ (Thomspon limit) 
where $\me$ is electron mass in energy units ;
both of these should
be valid for blazar emission
up to $\sim {100}\ \rm GeV$. 

\citet{Krawczynski:2011ym} revisited this calculation, both
analytically and numerically, 
and confirmed the results of  \citet{Bonometto1970}.
\citet{Krawczynski:2011ym} also computed, for the first time, 
the polarization from inverse-Compton scattering by an isotropic distribution of 
electrons in the Klein-Nishina regime. For a photon spectrum generated by an
isotropic distribution 
of electrons with a power-law energy spectrum of  index $p$
upscattering mono-energetic, unidirectional, 
$\Pi_{\rm init} = 100\%$ polarized 
photons, 
the resulting polarization fraction is \citep{Krawczynski:2011ym}
\beqar
\label{eq:pdegreeic}
\Pi_{e,\rm powerlaw} = \frac{(1+p)(3+p)}{(11+4p+p^{2})}\,.
\eeqar
The retained polarization is substantial:
for $p = 2 (3)$, $\Pi_{e,\rm powerlaw} \approx 65\%$ (75\%). These numbers 
assume the polarization and propagation 
directions of the polarized seed photons are perfectly aligned. 
In general, as 
shown in \citet{Krawczynski:2011ym}, this retained fraction 
depends on the polarization angle, the angle between 
the seed photon and the magnetic field direction, the spectral index, etc. 
Evaluating this complex geometric factor 
for each blazar is beyond the scope of this work. 
But in order to account for this effect, 
we make the simplifying assumption 
that the polarization and propagation direction of the seed 
photons are at a fixed average angle 
to retain a fraction, $\etaavg = 80\%$ of polarization. This is 
a reasonable number for blazar electron indices $\approx 2$ 
according to \citet{Krawczynski:2011ym}.
For an arbitrary initial polarization degree $\Pi_{\rm init}$,
the upscattered polarization  fraction 
becomes 
\beqar
\label{eq:krawczynksilaw}
\Pi_{\rm upscatter} = \etaavg \ \Pi_{e,\rm powerlaw} \ \Pi_{\rm init}\,
\eeqar

\subsection{Motivation for using observed polarization of low energy seed photons}
For blazars, the initial (``seed'') polarization $\Pi_{\rm init}$
is that of the low-energy photons, which themselves
arise from synchrotron emission.
Theoretically, for a power-law distribution 
of electrons, the degree of polarization for 
synchrotron is \citep[e.g.,][]{1986rpa..book.....R}, 
\beqar
\label{eq:syncpol}
\Pi_{\rm sync} = \frac{p+1}{p+7/3}\,.
\eeqar 
This would give large values, $\Pi_{\rm init} \approx 70\%$
for typical blazars where $p=2 - 3$. 
Optical observations show 
a high degree of polarization \citep[e.g.,][]{1980ARA&A..18..321A} 
always lower than this maximum synchrotron limit in several cases. 
This is because, as various effects like Faraday rotation 
and inhomogeneities in the magnetic field as also presence of 
thermal components can reduce this value \citep{2013peag.book.....N}.
Fermi-detected blazars with low degrees of polarization in the optical and IR have
been observed \citep{Fujiwara:2012gt, robopol}. In light of the diversity of the observed
low-energy blazar polarization,
we will restrict our predictions only to blazars with observed 
values of $\Pi_{\rm init}$. We thus propose to test empirically the 
relation between degrees of low 
energy polarization and high 
energy polarization.

In general, $\Pi$ is a function 
of the direction of target photons and 
energy of the scattered photons. Figures 
8 and 9 in \cite{Krawczynski:2011ym} show 
$\Pi$ to be an increasing function of the 
scattered photon energy, from energies of $\sim $100 {\rm keV}
to nearly 100 {\rm MeV}'s or so.
This suggests that soft gamma-rays are likely to be polarized. 
In general, this depends upon the target photons 
for the particular blazar type. 
However, a detailed calculation appropriate at 
hard X-ray / soft gamma-ray energies is needed to show this
explicitly. This is not the focus of this paper.

\section{Detection sensitivity of telescopes: Minimum Detectable Polarization}
\label{sect:sensitivity}

The detectability of polarization in a given source of certain 
strength by a given detector is quantified in terms 
of the minimum degree of polarization the detector can establish (the
minimum detectable polarization, MDP).  
The MDP is the minimum fraction polarized intensity detectable given the 
strength of the source, $R_{S}$ relative to the background, $R_{\rm bg}$ 
in a given exposure time, T.
 The MDP of a source producing 
a partially polarized signal, $R_{\rm S} \rct$,
is given by \citep{1975SSRv...18..389N, Weisskopf:2010se, Kalemci:2004p8246}, 
\beqar
\label{eq:mdp}
MDP = \frac{\nsig}{\mu\ R_{S}} \sqrt{\frac{2(R_{S} + R_{\rm bg})}{T} }\,,
\eeqar 
$\nsig$ is the detection significance, $R_{\rm bg}$ is the background
count rate, and $\mu$ is the modulation factor, which quantifies the
detector response to polarized light (with $\mu = 1$ being perfect
polarization sensitivity and $\mu=0$ being no polarization
sensitivity). Given a detector with a certain modulation factor $\mu$
and background $ R_{\rm bg} \rct $,
Eq.~(\ref{eq:mdp}) can be inverted to 
determine the amount of time 
required to detect a source producing $R_{S} \rct$. 

The observation time 
required to detect that a certain signal is polarized 
at a certain degree is different 
from the time required to make a measurement (both polarization degree
and angle) of 
its polarization with some given uncertainty. 
A measurement of the source's polarization properties constitutes 
the joint measurement of both the amplitude (degree, $\Pi$ ) 
and phase (related to the position angle, $\psi$) 
of polarization as opposed to simply the amplitude. This difference
is discussed 
by \citet{Weisskopf:2010se, 2013ApJ...773..103S} and the heart of
the issue is that more photons are needed in order 
to jointly measure both the degree and position angle, than to simply establish that
the source is polarized at a certain degree. Following the analysis in both \citet{Weisskopf:2010se, 2013ApJ...773..103S} and 
their suggested prescription, in this paper we 
make this distinction between the time 
required for a 3-sigma measurement 
of the MDP degree, $\Pi$ alone (establish polarization degree with only $1\%$
probability for the deviation from zero-polarization to be caused by chance) 
and the time required 
for a joint 3-sigma measurement of the degree, $\Pi$, corresponding to the MDP and 
position angle, $\psi$. We
tabulate both these times for the selected sources 
in section~\ref{sect:blazarlist}. The times differ by a 
factor $\approx \nsig^{2}/4 = 2.25$ for the polarization degrees of $\Pi \approx 10\%$.

The GEMS White Paper \citep{2013arXiv1303.7158K} 
proposes for GEMS-like missions a strategy 
similar to what we describe below. 
They emphasise using archival data 
on low energy polarization in order to motivate 
observations of high energy polarization. 
In this work we extend such calculations to polarimeters beyond GEMS-like
missions, and 
explicitly consider the effects of flux increase during blazar flaring.  
Based on this, we motivate the possibility of simultaneous polarization 
observations of blazars at low and high energies 
taking advantage of current and future 
low-energy blazar polarization monitoring programs, 
such as the RoboPol optopolarimetric program \citep{robopol}, and the
F-GAMMA\citep{fgamma}, and OVRO \citep{ovro} programs in radio frequencies.

\section{Detection prospects of blazars}
\label{sect:blazarlist}

\subsection{Selection of test cases}

We use blazars in \citet{2011PASJ...63..639I} and \citet{Fujiwara:2012gt} (selected from
the \fermi-LAT catalog and studied in IR polarization) as the pool
from which to select a few good candidates for high-energy
polarization detection and use them as demonstration cases in our
calculations below. Some of these are very bright, some possess a high degree 
of polarization, while others are variable and are good candidates 
for temporal studies. Thus, there are several alternate ways of 
selecting candidate blazars for polarimetric studies depending 
on both blazar characteristics and observability. 
3C 279, PKS 1510-089 and 
3C 454.3 are 3 of the brightest blazars that have the 
highest measured IR polarization recorded by various 
polarimeters as listed in \citet{2011PASJ...63..639I, Fujiwara:2012gt}. In addition,
they have shown significant variability \citep{2010Natur.463..919A, 2011PASJ...63..489S, 2012ApJ...758...72W, 2010ApJ...715..362J}. 
Their flaring 
behaviour make them promising candidates {\it a priori}
for polarization measurements as we will comment 
on in section~\ref{sect:flaring}. For these reasons, we will focus on
these three sources for the remainder of this section. 

\subsection{High-energy polarimeters}

Both main instruments on board INTEGRAL, IBIS and SPI, 
have reported polarization measurements of 
GRBs \citep[][etc.,]{2013MNRAS.431.3550G, 2009ApJ...695L.208G, 2007A&A...466..895M}. 
Thus, it is natural to 
check prospects of 
blazar measurements with INTEGRAL. In addition, there are a number of polarimeters in various stages of 
planning, commissioning and operation. We explore GEMS-like missions \citep{2013arXiv1303.7158K}
as
well as other instruments  \citep{2010xrp..book.....B} : balloon based 
polarimeters PoGOLITE \citep{2012arXiv1211.5094P} and X-Calibur \citep{2013APh....41...63G};
space based instruments ASTRO-H \citep{2012cosp...39.1259M} and POLARIX \citep{2010ExA....28..137C}, which are 
in the stage of planning; and the Gamma-Ray Burst 
Polarimeter (GAP) \citep{2011AIPC.1358..408Y}, which is 
already functional and has successfully detected 
polarized signals from some GRBs. 

GAP on board IKAROS is designed specifically for GRBs and was able 
to detect polarized signals from GRBs 110301A, 
110721A and 100826A \citep{2012ApJ...758L...1Y, 2011ApJ...743L..30Y}. 
Despite being suited for prompt polarimetry of GRBs, it is worth 
looking at prospects of blazar polarimetry with GAP, particularly 
those which have demonstrated high flux variations like the 3 we consider.

\subsection{Strategy of calculations}

We use observed polarization values, $\Pi_{\rm init}$
to determine the degree of high energy polarization of a blazar, $\Pi_{\rm upscatter,blazar}$
as described in section~\ref{sect:degree}. In order to do this, we scale 
from the sensitivity of polarimeters to a Crab-like source 
for each blazar. We denote this source using the subscript ``calib'' (for calibrator), as 
this is essentially a way of calibrating the response 
of the polarimeter. The sensitivity of polarimeters 
are given in terms of the amount of time taken 
to detect the MDP from a source of a given flux (typically expressed 
in mCrab units) in accordance with Eq.~\ref{eq:mdp}. 

Scaling from the Crab using Eq.~(\ref{eq:mdp}), 
we get then the required observation times, $\tobs$. Here there are 3 cases, if the 
background for the polarization signal is same for 
both the calibrator and blazar, noting that this may not 
always hold true in practice. First, 
when the source signal is much higher 
than the background for both the calibrator source and the blazar i.e. 
$\rbl, \rcrab \gg R_{\rm bg}$. Second, 
when the background dominates 
over the signal for both i.e.,$\rbl, \rcrab \ll R_{\rm bg}$. Third, when the 
calibrator signal dominates the background, 
but the background dominates the blazar signal i.e., $\rcrab \gg R_{\rm bg} \gg \rbl$. 
A fourth case would occur if the blazar signal dominates 
the background, but the calibrator doesn't. However, 
it is highly unlikely that the blazars are brighter 
than the calibrator source which is typically the Crab. 

In the first case, when the blazar and calibrator count rates are much higher 
than the background i.e. $\rcrab,\rbl \gg R_{\rm bg}$, the required observation
times are given by, 
\beqar
\label{eq:sourcedomobstime}
\frac{\tobsbl}{\tobscrab}\bigg|_{\rm sig\ dom} = \pfrac{\rcrab}{\rbl} \pfrac{\mdpcrab}{\Pi_{\rm upscatter, blazar}}^{2} 
\eeqar
In the second case, when both the blazar and calibrator count rates 
are much lower than the background, i.e.,$\rcrab,\rbl \ll R_{\rm bg}$, the required observation
times are given by,
\beqar
\label{eq:bgddomobstime}
\frac{\tobsbl}{\tobscrab}\bigg|_{\rm bgd\ dom} = \pfrac{\rcrab}{\rbl}^{2} \pfrac{\mdpcrab}{\Pi_{\rm upscatter, blazar}}^{2}
\eeqar
respectively. This is true provided the background is the 
same for the calibrator and the blazar such that it drops out from the ratio 
of observing times. 
In the third case i.e., $\rcrab \gg R_{\rm bg} \gg \rbl$, 
the observation time is given by, 
\beqar
\label{eq:bgddomblazarobstime}
\frac{\tobsblbgd}{\tobsblcalsig} = \pfrac{\rcrab\ R_{\rm bg}}{\rbl^{2}} \pfrac{\mdpcrab}{\Pi_{\rm upscatter, blazar}}^{2} \nn
\eeqar
From the above it is clear that the third case is 
gives an intermediate value for required observation time 
compared to the first two cases, 
as $\frac{\rcrab}{\rbl} \gg \frac{R_{\rm bg}}{\rbl} \gg 1$ in 
the third case. 
The actual observation time will 
be within the range set by these first two limiting cases given 
by eqns.~\ref{eq:sourcedomobstime} and ~\ref{eq:bgddomobstime}, that 
are independent of the background count rate.  
This is of course true if the background rates 
for the calibrator and blazar are the same or approximately the same. 
If the calibrator and blazar backgrounds are different, 
their ratio would appear in the equation~\ref{eq:bgddomobstime}. 
This would increase the maximum required observing 
time in the case that the background rate for the blazar is much higher 
than that for the calibrator source. In this case, the polarization 
signal from the blazar would 
simply be very hard to observe. 

Here, we assume that 
the degree of high energy polarization of 
a blazar due to inverse-Compton 
scattering calculated from eqn.~\ref{eq:krawczynksilaw} 
should be measurable 
by the polarimeter. Therefore, we determine 
time required to do so given a dedicated 
observation of the particular blazar.  We drop the subscript ``blazar'' 
in the tables.  

\subsection{Inputs }
\subsubsection{Flux of the Crab in different energy ranges}
\label{sect:Crab}
\begin{table*}[ht!]
\begin{center}
\begin{tabular}{|c|c|c|c|c|c|c|c|}
\hline
\hline
Polarimeter energy range & Observed Energy Range & Photon
& Value of mCrab & Value of mCrab &  Value of mCrab & Calibrator flux \\ 
$\delepoli{j} = $& $\deleobsi{j} = $ & index & Observed energy flux & Scaled energy flux& Scaled photon flux & \\
$\epolmini{j}-\epolmaxi{j}$ & $\eobsmini{j}-\eobsmaxi{j}$ & $\G{Crab}{j}$ & $\feobscali{j}$ & $\fepolcali{j}$& $\fphpolcali{j}$ & \\
(keV) & (keV) & &($10^{-12}{\rm ergs\ cm^{-2}\ sec^{-1}}$) & ($10^{-12}{\rm ergs\ cm^{-2}\ sec^{-1}}$) & ($10^{-5}{\rm \ cm^{-2}\ sec^{-1}}$) & (mCrab) \\
 \hline
PoGOLITE : $25-80$ & $17-60$&  2.1 & $14.3$ & $12.8$ & $19.1$ & 200\\
\hline
X-Calibur : $25-80$ & $17-60$ & 2.1 & $14.3$ & $12.8$ & $19.1$ & 25 \\
\hline
GEMS-like :  $2-10$ & $2-10$ & 2.1 & $2.4\times10^{1}$ & $2.4\times10^{1}$ & $3.8\times10^{2}$ & 1\\
\hline
POLARIX : $2-10$ & $2-10$ & 2.1 & $2.4\times10^{1}$ & $2.4\times10^{1}$ & $3.8\times10^{2}$ & 1 \\
\hline
ASTRO-H : $50-195$ & $14-195$ & 2.1& $23.86$ & $9.151$ & $6.344$ & 10 \\
\hline
GAP : $50-195$ & $14-195$ & 2.1& $23.86$ & $9.151$ & $6.344$& 1000\\
\hline
\end{tabular}
\caption{Values of mCrab for the energy range of polarimeters are listed here and the way they are obtained 
for the appropriate energy range from existing data is described section~\ref{sect:Crab}. We list the energy fluxes in observed energy ranges and the scaled energy fluxes appropriate for the polarimeter energy range, $\delepoli{j} = \epolmini{j} - \epolmaxi{j}$, using the spectral index, $\G{calib}{j}$ of the Crab in the observed energy range, $\eobsmini{j}-\eobsmaxi{j}$. In the final column, we also list the calibrator flux values in mCrab.}
\label{table:mCrabvalues}
\end{center}
\end{table*}
In order to scale off the calibrator source, either we need to 
express blazar fluxes in units of
mCrab or convert the flux of the calibrator and blazar in 
the same units (${\rm photons\ cm^{-2}\ sec^{-1}}$ is used).  Because of the difference between the Crab energy spectrum and
the candidate-blazar spectra, this conversion is energy-dependent, and
so our first input is the flux of the Crab in each instrument's energy
range. In order to determine the photon number 
flux of the calibrator source, there are two steps. First, we 
derive the energy flux in the
polarimeter range for each of the 6 instruments, $\fepolcali{j}$ from the observed energy flux $\feobscali{j}$ listed in column 4 of Table~\ref{table:mCrabvalues}, 
where $j = 1-6$ corresponding to the polarimeter index. Here, $\delepoli{j} = \epolmini{j}-\epolmaxi{j}$ 
symbolizes the energy range of 
the polarimeter considered and $\deleobsi{j}=\eobsmini{j}-\eobsmaxi{j}$ symbolizes 
the energy range of observed fluxes used as input. 
And then we convert this scaled energy flux at 
the polarimeter energies, $\fepolcali{j}$ or column 5 of Table~\ref{table:mCrabvalues} into a photon flux, $\fphpolcali{j}$ 
in the same energy range listed in column 6. The final column 7, 
lists the calibrator flux values from column 6 in mCrab units. Since, we use observed energy fluxes 
as input, we chose the observed values in the energy range closest 
to the polarimeter range and use the photon index $\Gamma$ to scale.  
For the 6 polarimeters under consideration, the energy 
ranges are labelled as $(\epolmini{j}-\epolmaxi{j})$. PoGOLITE 
and X-Calibur have $\epolmini{1,2} = 25$ keV and $\epolmaxi{1,2} = 80$ keV, 
GEMS-like instruments 
and POLARIX have $\epolmini{3,4} = 2$ keV and $\epolmaxi{3,4} = 10$ keV, 
and finally ASTRO-H and GAP have $\epolmini{5,6} = 50$ keV and $\epolmaxi{5,6} = 195$ keV
as indicated in Table~\ref{table:mCrabvalues}. 

IBIS / INTEGRAL  measured the Crab flux in 
the energy range from $\eobsmini{1} - \eobsmaxi{1} = 17 - 60$ keV 
{\label{CrabIBISflux}\citep{2010A&A...519A.107K}}\footnote{\url{http://heasarc.gsfc.nasa.gov/db-perl/W3Browse/w3table.pl?tablehead=name\%3Dintibisass&Action=More+Options}}. In order to scale
to the energy range of 
PoGOLITE and X-Calibur, we use this flux 
and scale with the 
spectral index from SPI / INTEGRAL {\label{CrabIBISspecindex}{\citep{2008int..workE.144J}}. 
For GEMS-like missions 
and for POLARIX we use the Crab flux
measured by RXTE in the same 
energy range, $\Delta E_{\rm pol} = \Delta E_{\rm obs} = 2-10$ keV , 
\citep{2010ApJ...713..912W}. 
This scaling is done by the following equation,  
\beqar
\label{eq:Crabobstopolenergyband}
\hspace*{-10cm}\fepolcali{j} = \feobscali{j} \hspace*{1.5cm} \\  
\times \frac{\epolmini{j}^{2-\G{calib}{j}} - \epolmaxi{j}^{2-\G{calib}{j}} }{\eobsmini{j}^{2-\G{calib}{j}} - \eobsmaxi{j}^{2-\G{calib}{j}}} \nn  
\eeqar 
for $j = 1-4$. $\G{calib}{j}$ is the photon index of the Crab in the 
considered range of energy. For $j = 3,4$, the eqn.~\ref{eq:Crabobstopolenergyband} 
trivially holds true.

For ASTRO-H and GAP ( $\Delta E_{\rm pol} = 50 - 195$ keV)\footnote{ In practice,  
ASTRO-H is expected to have polarimetry information 
from 50 - 300 keV, but here we will consider the range below 195 keV where observed flux  
values are available.}, we
use here  the Crab flux in the energy range $\Delta E_{\rm obs} = 14 - 195$ keV 
from the 70 Month Swift - BAT All - sky Hard X - Ray Survey 
\citep{2013ApJS..207...19B} and an average  
spectral index of 2.1 from several measurements \citep{2005SPIE.5898...22K}. 
Then the Crab flux in the energy range from 
between the minimum energy of the observed range $\eobsmini{5,6} = 14$ keV 
and the minimum polarimeter energy $\epolmini{5,6} = 50$ keV is 
derived from the IBIS flux between $\eobsmini{1} - \eobsmaxi{1} = 17 - 60$ keV. 
This flux is subtracted from BAT band flux, $\feobscali{1}$ to 
give the calibrator flux in the ASTRO-H / GAP energy range as, 
\beqar
\label{eq:Crabobstopolenergybandastrohgap}
\fepolcali{j} &=& \feobscali{j} - \feobscali{1} \\
\! \! \! \! \! &\times& \frac{\eobsmini{j}^{2-\G{calib}{1}} - \epolmini{j}^{2-\G{calib}{1}} }{\eobsmini{1}^{2-\G{calib}{1}} - \eobsmaxi{1}^{2-\G{calib}{1}}} \nn
\eeqar
for $j = 5,6$. Note that the numerator of the second term in 
the above equation, correctly has $\eobsmini{j}$ and 
$\epolmini{j}$ as we want to subtract the flux in the range from 
$\eobsmini{j} - \epolmini{j} = 14 - 50$ keV. For the Crab, since the spectral index is 
$\approx$ 2.1 for both the IBIS and the BAT range one can simply use this value 
for both the observed and polarimeter energy ranges. 
For blazars, the same approach 
is followed with the energy range for which 
observed fluxes are available.

After determining the energy flux in the polarimeter energy 
range, the second step is to calculate the photon number flux as
\beqar
\label{eq:energytophotonfluxcal}
\fphpolcali{j} &=& \frac{\fepolcali{j} }{\epolmini{j}} \frac{2-\G{calib}{j}}{1-\G{calib}{j}} \\
&\times& \frac{1 -\pfrac{\epolmaxi{j}}{\epolmini{j}}^{1-\G{calib}{j}} }{1 -\pfrac{\epolmaxi{j}}{\epolmini{j}}^{2-\G{calib}{j}}} \ \ \ \ \ \nn
\eeqar
for $j = 1-6$.
The Crab fluxes
(both energy fluxes and photon counts)
in the various energy ranges of the instruments 
of interest are tabulated in table~\ref{table:mCrabvalues}. 

\subsubsection{Fluxes, photon, and electron indices, seed polarization of blazars}
\label{sect:blazarparameters}
\begin{table*}[ht!]
\begin{tabular}{|c|c|c|c|c|c|c|c|}
\hline
\hline
Polarimeter id & Polarimeter  & Energy range  & Observed energy
& Scaled energy & Photon index & Electron index &
 Seed \\
j & energy range & of observed flux & flux of blazar  & flux of blazar & of blazar & of blazar &Polariz. degree\\
& $\delepoli{j}$ & $\deleobsi{j}$ &$\feobsbli{j}$ & $\fepolbli{j}$ &  $\G{blazar}{j}$& p(j) & $\Pi_{\rm init}$ \\
 & (keV) &(keV)  & ($\times10^{-12}{\rm ergs\ cm^{-2}\ sec^{-1}}$)& ($\times10^{-12}{\rm ergs\ cm^{-2} \ sec^{-1}}$)&  & &  ($\%$) \\ 
 \hline
1,2 & 25 - 80 & 17 - 60 & $12.40$& $13.07$& 1.6 & 2.2 & 20.0 \\
\hline
3,4 & 2 - 10 &  2 - 10 & $10.00$ &  $10.00$ & 1.56 & 2.12 & 20.0 \\
\hline
5,6 & 50 - 195 & 14 - 195 & $34.3$&  $22.7$ & 1.49 &1.98 & 20.0 \\
\hline
\end{tabular}
\caption{Input parameters for 3C 279 with detailed description in section~\ref{sect:blazarparameters}. We list the energy fluxes in observed energy ranges and the scaled energy fluxes appropriate for the polarimeter energy range, $\delepoli{j} = \epolmini{j} - \epolmaxi{j}$, using the spectral index, $\Gamma$ of 3C 279 in the observed energy range, $\eobsmini{j}-\eobsmaxi{j}$. In the final column, we also list the seed polarization degree value (temporal average). For ASTRO-H and GAP, we scale fluxes to BAT energy range of 50-195.}
\label{table:3C279input}
\end{table*}

\begin{table*}[ht!]
\begin{tabular}{|c|c|c|c|c|c|c|c|}
\hline
\hline
Polarimeter id &Polarimeter  & Energy range  & Observed energy
 & Scaled energy & Photon index & Electron index &
 Seed \\
j & energy range & of observed flux & flux of blazar  & flux of blazar & of blazar & of blazar &Polariz. degree\\
& $\delepoli{j}$ & $\deleobsi{j}$ &$\feobsbli{j}$ & $\fepolbli{j}$ &  $\G{blazar}{j}$& p(j) & $\Pi_{\rm init}$ \\
 & (keV) &(keV)  & ($\times10^{-12}{\rm ergs\ cm^{-2}\ sec^{-1}}$)& ($\times10^{-12}{\rm ergs\ cm^{-2} \ sec^{-1}}$)&  & &  ($\%$) \\ 
 \hline
 1,2 & 25 - 80 & 10 - 50 &$38.2$ & $45.7$& 1.23 & 1.46 & 15.0 \\
 \hline
 3,4 & 2 - 10 &  2 - 10 &$6.09$ & $6.09$ & 1.38 & 1.76 &15.0  \\
\hline
5,6 & 50 - 195 & 14 - 195 &$70.0$& $36.4$& 1.38 & 1.76 & 15.0  \\
\hline
\end{tabular}
\caption{Input parameters for PKS 1510-089 with detailed description in section~\ref{sect:blazarparameters} as in Table~\ref{table:3C279input} for 3C 279.}
\label{table:PKS1510input}
\end{table*}

\begin{table*}[ht!]
\begin{tabular}{|c|c|c|c|c|c|c|c|}
\hline
\hline
Polarimeter id & Polarimeter  & Energy range  & Observed energy
 & Scaled energy & Photon index & Electron index &
 Seed \\
j & energy range & of observed flux & flux of blazar  & flux of blazar & of blazar & of blazar &Polariz. degree\\
& $\delepoli{j}$ & $\deleobsi{j}$ &$\feobsbli{j}$ & $\fepolbli{j}$ &  $\G{blazar}{j}$& p(j) & $\Pi_{\rm init}$ \\
& (keV) &(keV)  & ($\times10^{-12}{\rm ergs\ cm^{-2}\ sec^{-1}}$)& ($\times10^{-12}{\rm ergs\ cm^{-2} \ sec^{-1}}$)&  & &  ($\%$) \\ 
 \hline
1,2 & 25 - 80 & 17 - 60 & $1.03\times10^{2}$ & $1.02\times10^{2}$&1.8 & 2.6 &10.0\\
\hline
3,4 & 2 - 10 &  2 - 10 &$17.6$ & $17.6$ & 1.53 & 2.06 & 10.0 \\
\hline
5,6 & 50 - 195 & 14 - 195 & $133.0$ &  $32.87$ & 1.52 & 2.04 & 10.0 \\
\hline
\end{tabular}
\caption{Input parameters for 3C 454.3 with detailed description in section~\ref{sect:blazarparameters} as in Table~\ref{table:3C279input} for 3C 279.}
\label{table:3C454input}
\end{table*}

The photon fluxes of the selected blazars are 
computed just like for the Crab as described in the previous section. 
First the energy fluxes in observed ranges, 
are scaled to the polarimeter energy ranges with 
the appropriate spectral indices $\G{blazar}{j}$, as
\beqar
\label{eq:blazarobstopolenergyband}
\hspace*{-10cm}\fepolbli{j} = \feobsbli{j}  \hspace*{2cm} \\ 
\times \frac{\epolmini{j}^{2-\G{blazar}{j}} - \epolmaxi{j}^{2-\G{blazar}{j}} }{\eobsmini{j}^{2-\G{blazar}{j}} - \eobsmaxi{j}^{2-\G{blazar}{j}}} \nn\ \ \ \ \ \ 
\eeqar 
for $j = 1-4$. 
And as in the previous section 
describing the calibrator flux calculation, for ASTRO-H and GAP, the 
energy fluxes appropriate 
for the polarimeter ranges 
need to be calculated with 
\beqar
\label{eq:Crabobstopolenergybandastrohgap}
\hspace*{-0.5cm} \fepolbli{j} = \feobsbli{j} - \feobsbli{1} \hspace*{0.6cm} \\
\hspace*{-1.cm} \times \frac{\eobsmini{j}^{2-\G{blazar}{1}} - \epolmini{j}^{2-\G{blazar}{1}} }{\eobsmini{1}^{2-\G{blazar}{1}} - \eobsmaxi{1}^{2-\G{blazar}{1}}} \nn
\eeqar
for $j = 5,6$.
The fluxes and the photon indices 
of the blazars are taken from
the literature. From 
the photon index, $\Gamma$, the electron 
index is calculated in the Thomson limit in 
the electron rest frame, $p = 2\Gamma -1$. 
This is used to calculate the  
high energy polarization fraction due to IC 
scattering, $\Pi_{e,\rm powerlaw} $ 
from eqn.~\ref{eq:pdegreeic}. 

For 3C 279, the IBIS flux from the 
INTEGRAL IBIS All Sky Survey (INTIBISASS) 
of Hard X-ray Sources 
\footnote{\url{http://heasarc.gsfc.nasa.gov/db-perl/W3Browse/w3table.pl?tablehead=name\%3Dintibisass&Action=More+Options}}
\citep{2010A&A...523A..61K} is used 
and scaled to PoGOLITE and X-Calibur 
energy ranges using the spectral index 
from Swift BAT AGN 60 month survey \citep{2012ApJ...749...21A}. 
For GEMS-like and POLARIX, the mean RXTE flux from figure 4 
in \citep{2010A&A...522A..66C} 
that is consistent with Swift XRT measurements is used. And spectral 
index is the Swift XRT value in \citep{2010A&A...522A..66C}. 
For ASTRO-H and GAP, 
the Swift BAT 70 month survey fluxes \citep{2013ApJS..207...19B} 
and Swift BAT 60 month 
AGN catalog indices \citep{2012ApJ...749...21A} 
\footnote{\url{http://heasarc.gsfc.nasa.gov/db-perl/W3Browse/w3table.pl?tablehead=name\%3Dswbatagn60&Action=More+Options}}
are used in the range $14 - 195$ keV. The IBIS flux 
for 3C 279 in the 17 - 60 keV band is then scaled to 14 - 50 keV and 
subtracted from the BAT flux to match the appropriate range. 

Similarly, fluxes and indices are determined for 
PKS 1510-089 and 3C 454.3. For 
PKS 1510-089, the fluxes taken are
Suzaku \citep{2008ApJ...672..787K} (10 - 50 keV) for PoGOLITE and X-Calibur, Swift (2 - 10 keV) \citep{2010ApJ...716..835A}
for GEMS-like and POLARIX, and the BAT 70 month 
survey fluxes for ASTRO-H and GAP \citep{2013ApJS..207...19B}. Both the photon 
index and flux values for PKS 1510-089 
are the Suzaku values in table~\ref{table:PKS1510input} for the power-law + disk blackbody model ($PL+DB$) model of \citet{2008ApJ...672..787K} 
in the $10-50$ keV range. And for $14-195$ keV, the photon index along with the 
flux is from Swift-BAT 70 month Hard X-ray survey \citep{2013ApJS..207...19B}  
Like for PKS 1510-089, for 
3C 454.3 the fluxes are from 
INTIBISASS, Swift-XRT and BAT respectively. 
The photon index for 3C 454.3 to scale to the $25-80$ keV range 
is taken from combined analysis of IBIS/ISGRI, SPI and JEM-X data done in \citet{2006A&A...449L..21P}. 
For $2-10$ keV, the values are from 
are from Swift-XRT \citep{2010ApJ...716..835A} for $2-10$ keV   
and for $14-195$ keV, again Swift-BAT is used \citep{2013ApJS..207...19B}. 
Once, we get the energy fluxes in the polarimeter energy ranges, 
we can readily compute the photon fluxes using the spectral indices as 
was done for the calibrator, 
\beqar
\label{eq:energytophotonfluxblazar}
\fphpolbli{j} &=& \frac{\fepolbli{j} }{\epolmini{j}} \frac{2-\G{blazar}{j}}{1-\G{blazar}{j}} \nn\\
&\times&\frac{1 -\pfrac{\epolmaxi{j}}{\epolmini{j}}^{1-\G{blazar}{j}} }{1 -\pfrac{\epolmaxi{j}}{\epolmini{j}}^{2-\G{blazar}{j}}} \ \ \ \ 
\eeqar
We use these photon fluxes, $\fphpol$ as inputs in eqns.~(\ref{eq:sourcedomobstime}) and 
~(\ref{eq:bgddomobstime}) instead of the source 
count rates, $R$. In absence of detailed energy dependent models of 
effective areas for every polarimeter, a consistent 
way of determine the photon count rates for every blazar is difficult. And 
hence, ignoring distinction between $\fphpol$ and R we put
\beqar
\label{eq:rbleqfphbl}
\! \! \! \! \! \! \rbl = \fphpolbl \ ; \ \rcrab = \fphpolcal
\eeqar
This difference is only important for those blazars 
whose photon indices are different from the average Crab index of 
2.1. An approximate calculation for ASTRO-H 
assuming a power-law dependence of effective area 
on energy 
\footnote{\url{http://astro-h.isas.jaxa.jp/researchers/sim/effective\_area.html}}
shows that for PKS 1510-089 with the most different spectral index of 
$\approx 1.4$, there is approximately a factor of 2 difference 
in the ratio of $\rbl / \rcrab$ when the energy dependence 
of the effective area is accounted for. Thus, it is not a significant difference 
in the observation times relative to 
the range calculated here. 

Seed photon IR polarization values are taken from average 
values consistent with \citet{Fujiwara:2012gt, 2011PASJ...63..639I}. 
All these inputs are listed in Tables 
~(\ref{table:3C279input}), ~(\ref{table:PKS1510input}), ~(\ref{table:3C454input}). 

\begin{table*}[ht!]
\rotatebox{0}{
\scalebox{1.0}{
\begin{tabular}{|c|c|c|c|c|c|c|}
\hline
\hline
Source & Photon Flux & MDP & Photon Flux & High Energy  &  3-$\sigma$ $\Pi$ & 3-$\sigma$ ($\Pi$, $\psi$)\\
& Sensitivity ($\rcrab$) & & ($\rbl$) & Polariz.~Degree &  measurement
time& measurement time \\
 & (${\rm \times 10^{-5}\ cm^{-2}\ sec^{-1}}$) & $(\%)$& (${\rm \times 10^{-5}\ cm^{-2}\ sec^{-1}}$)& $\Pi_{\rm upscatter}$$(\%)$ & $\frac{\tobsbl}{\tobscrab}$ & $\frac{\nsig^{2}}{4} \frac{\tobsbl}{\tobscrab}$  \\
\hline
3C 279 & $3.82\times10^{3}$  & 10.0 & $18.50$ & 10.81 &$1.77\times10^{2} - 3.65\times10^{4}$ & $3.98\times10^{2} - 8.21\times10^{4}$\\
\hline
PKS 1510-089 & & & $62.0$ & 6.94 &$(1.28 - 78.8)\times10^{2}$ & $2.88\times10^{2} - 1.78\times10^{4}$\\
\hline
3C 454.3 & & & $1.47\times10^{2}$ & 5.73 &$7.92\times10^{1} - 2.06\times10^{2}$ & $1.78\times10^{2} - 4.63\times10^{3}$\\
\hline
\end{tabular}
}
}
\caption{Observational prospects for PoGOLITE (25 - 80 \rm keV): 200 mCrab source at $10\%$ \\
polarization detectable in 1 flight $(20\  days)$. Measurement times calculated using eqns~\ref{eq:sourcedomobstime} and ~\ref{eq:bgddomobstime}.}
\label{table:PoGOLITE}
\end{table*}

\begin{table*}[ht!]
\rotatebox{0}{
\scalebox{1.0}{
\begin{tabular}{|c|c|c|c|c|c|c|}
\hline
\hline
Source & Photon Flux & MDP & Photon Flux & High Energy  &  3-$\sigma$ $\Pi$ & 3-$\sigma$ ($\Pi$, $\psi$) \\
& Sensitivity ($\rcrab$) & & ($\rbl$) & Polariz.~Degree & measurement
time& measurement time \\
 & (${\rm \times 10^{-5}\ cm^{-2}\ sec^{-1}}$) & $(\%)$& (${\rm \times 10^{-5}\ cm^{-2}\ sec^{-1}}$)& $\Pi_{\rm upscatter}$$(\%)$ & $\frac{\tobsbl}{\tobscrab}$ & $\frac{\nsig^{2}}{4} \frac{\tobsbl}{\tobscrab}$  \\
\hline
3C 279 & $4.78\times10^{2}$  & 4.5 & $18.50$ & 10.81 & $4.47 - 115.39$ & $10.07 - 2.60\times10^{2}$\\
\hline
PKS 1510-089 & & & $62.0$ & 6.94 &$3.24 - 25.0$ & $7.29 - 56.2$\\
\hline
3C 454.3 & & & $1.47\times10^{2}$ & 5.73 & $2.00 - 6.52$ & $4.51-14.7$\\
\hline
\end{tabular}
}
}
\caption{Observational prospects for X-Calibur (25 - 80 \rm keV): 25 mCrab source at $4.5\%$ \\ polarization is detectable in 1000 ksec. Measurement times calculated using eqns~\ref{eq:sourcedomobstime} and ~\ref{eq:bgddomobstime}.}
\label{table:X-Calibur}
\end{table*}

\begin{table*}[ht!]
\rotatebox{0}{
\scalebox{1.0}{
\begin{tabular}{|c|c|c|c|c|c|c|}
\hline
\hline
Source & Photon Flux & MDP & Photon Flux & High Energy  &  3-$\sigma$ $\Pi$ & 3-$\sigma$ ($\Pi$, $\psi$) \\
& Sensitivity ($\rcrab$) & & ($\rbl$) & Polariz.~Degree & measurement
time& measurement time \\
 & (${\rm \times10^{-5}\ cm^{-2}\ sec^{-1}}$) & $(\%)$& (${\rm \times10^{-5}\ cm^{-2}\ sec^{-1}}$)& $\Pi_{\rm upscatter}$$(\%)$ & $\frac{\tobsbl}{\tobscrab}$ & $\frac{\nsig^{2}}{4} \frac{\tobsbl}{\tobscrab}$  \\
\hline
3C 279 & $3.8\times10^{2}$  & 2.0 & $142.0$ & 10.66 &$9.4\times10^{-2} - 2.5\times10^{-1}$ & $(2.1-5.7)\times10^{-1}$\\
\hline
PKS 1510-089 & & & $83.0$ & 7.46 &$0.3-1.5$ & $7.4\times10^{-1} - 3.4$\\
\hline
3C 454.3 & & & $2.48\times10^{2}$ & 5.28 &$2.2\times10^{-1} - 3.4\times10^{-1}$ & $(5.0-7.6)\times10^{-1}$\\
\hline
\end{tabular}
}
}
\caption{Observational prospects for GEMS-like instrument (2 - 10 keV): 1 mCrab source \\ at $2 \%$ polarization is detectable in 1000 ksec. Measurement times calculated using eqns~\ref{eq:sourcedomobstime} and ~\ref{eq:bgddomobstime}.}
\label{table:GEMS}
\end{table*}

\begin{table*}[ht!]
\rotatebox{0}{
\scalebox{1.0}{
\begin{tabular}{|c|c|c|c|c|c|c|}
\hline
\hline
Source & Photon Flux & MDP & Photon Flux & High Energy  & 3-$\sigma$ $\Pi$ & 3-$\sigma$ ($\Pi$, $\psi$) \\
& Sensitivity ($\rcrab$) & & ($\rbl$) & Polariz.~Degree & measurement
time& measurement time \\
 & (${\rm \times10^{-5}\ cm^{-2}\ sec^{-1}}$) & $(\%)$& (${\rm \times10^{-5}\ cm^{-2}\ sec^{-1}}$)& $\Pi_{\rm upscatter}$$(\%)$ & $\frac{\tobsbl}{\tobscrab}$ & $\frac{\nsig^{2}}{4} \frac{\tobsbl}{\tobscrab}$  \\
\hline
3C 279 & $3.8\times10^{2}$  & 10.0 & $142.0$ & 10.66 &$2.4 - 6.3$ & $5.3\times10^{0} - 1.4\times10^{1}$\\
\hline
PKS 1510-089 & & & $83.0$ & 7.46 &$8.2\times10^{0} - 3.8\times10^{1}$ & $(1.9 - 8.5)\times10^{1}$\\
\hline
3C 454.3 & & & $2.48\times10^{2}$ &  5.28 &$5.5 - 8.4$ & $1.2\times10^{1} - 1.9\times10^{1}$\\
\hline
\end{tabular}
}
}
\caption{Observation prospects for POLARIX (2.0-10.0 keV): 1 mCrab source at \\ $10 \%$ polarization is detectable in 100 ksec. Measurement times calculated using eqns~\ref{eq:sourcedomobstime} and ~\ref{eq:bgddomobstime}.}
\label{table:POLARIX}
\end{table*}

\begin{table*}[ht!]
\rotatebox{0}{
\scalebox{1.0}{
\begin{tabular}{|c|c|c|c|c|c|c|}
\hline
\hline
Source & Photon Flux & MDP & Photon Flux & High Energy  &  3-$\sigma$ $\Pi$ & 3-$\sigma$ ($\Pi$, $\psi$) \\
& Sensitivity ($\rcrab$) & & ($\rbl$) & Polariz.~Degree & measurement
time& measurement time \\
 & (${\rm \times10^{-5}\ cm^{-2}\ sec^{-1}}$) & $(\%)$& (${\rm \times10^{-5}\ cm^{-2}\ sec^{-1}}$)& $\Pi_{\rm upscatter}$$(\%)$ & $\frac{\tobsbl}{\tobscrab}$ & $\frac{\nsig^{2}}{4} \frac{\tobsbl}{\tobscrab}$ \\
\hline
3C 279 & $6.344\times10^{1}$  & 4.3 & $14.3$ & 10.40&$7.58\times10^{-1} - 3.37$ & $1.71 - 7.57$\\
\hline
PKS 1510-089 & & & $22.6$ & 7.46 &$0.93 - 2.62$ & $2.10 - 5.89$\\
\hline
3C 454.3 & & & $20.61$ &  5.26 &$2.06 - 6.33$ & $4.59  - 14.0$\\
\hline
\end{tabular}
}
}
\caption{Observational prospects for ASTRO-H (50-195 keV): 10 mCrab \\ source at $4.3 \%$ is detectable in 1000 ksec. Measurement times Calculated using eqns~\ref{eq:sourcedomobstime} and ~\ref{eq:bgddomobstime}.}
\label{table:ASTRO-H}
\end{table*}

\begin{table*}[ht!]
\rotatebox{0}{
\scalebox{1.0}{
\begin{tabular}{|c|c|c|c|c|c|c|}
\hline
\hline
Source & Photon Flux & MDP & Photon Flux & High Energy  &  3-$\sigma$ $\Pi$ & 3-$\sigma$ ($\Pi$, $\psi$) \\
& Sensitivity ($\rcrab$) & & ($\rbl$) & Polariz.~Degree & measurement
time& measurement time \\
 & (${\rm \times10^{-5}\ cm^{-2}\ sec^{-1}}$) & $(\%)$& (${\rm \times10^{-5}\ cm^{-2}\ sec^{-1}}$)& $\Pi_{\rm upscatter}$$(\%)$ & $\frac{\tobsbl}{\tobscrab}$ & $\frac{\nsig^{2}}{4} \frac{\tobsbl}{\tobscrab}$  \\
\hline
3C 279 & $6.344\times10^{3}$  & 20.0 & $14.3$ & 10.40& $1.64\times10^{3} - 7.28 \times10^{5}$ & $3.69\times10^{3} - 1.64\times10^{6}$\\
\hline
PKS 1510-089 & & & $22.6$ & 7.46 &$2.02\times10^{3}-5.66\times10^{5}$ & $4.54\times10^{3}- 1.27\times10^{6}$\\
\hline
3C 454.3 & & & $20.61$ &  5.26 &$4.41\times10^{3} - 1.37\times10^{6}$ & $1.00\times10^{4}  - 3.03\times10^{6}$\\
\hline
\end{tabular}
}
}
\caption{Observational prospects for GAP (50.0-300.0 keV): 1 Crab source at $20 \%$ \\ polarization is detectable in 2 days.  Measurement times calculated using eqns~\ref{eq:sourcedomobstime} and ~\ref{eq:bgddomobstime}.}
\label{table:GAP}
\end{table*}

\subsection{Polarimeter sensitivities $\&$ polarization detection prospects} 
\subsubsection{Detection times for the polarimeters}
The polarimeter sensitivities
are expressed in terms of their ability to detect polarization 
from a Crab-like signal quantified in terms 
of the minimum detectable polarization in Eq.~(\ref{eq:mdp}). 
The high-energy polarization degree for each test source is 
derived from Eq.~\ref{eq:pdegreeic} using 
seed photon IR polarization values from \citet{2011PASJ...63..639I, Fujiwara:2012gt}, and,
together with the other inputs discussed above, can be 
used to compute the exposure or observation time, $\tobsbl$ 
required to detect the polarization degree from the blazar under consideration. 

As an illustration we look at the case 
of polarimetry with INTEGRAL. 
Both main instruments on board INTEGRAL, IBIS and SPI, 
have reported polarization measurements of 
GRBs \citep[][etc.,]{2013MNRAS.431.3550G, 2009ApJ...695L.208G, 2007A&A...466..895M}. 
Thus, it is natural to 
check prospects of 
blazar measurements with INTEGRAL. 
With a modulation factor of $\sim 20 \%$ \cite{Kalemci:2004p8246} 
in the spectrometer on INTEGRAL, SPI, for the Crab, 
with $R_{S} = 0.18 \rct$ and $R_{\rm bg} = 14 \rct$, 
an exposure of 1000 ksec gives $MDP =44.4\ \%$. 
3C454.3, one of the brightest blazars 
which is also known to flare, has a 
flux of $1.85\times10^{-10} \reflux$ \citep{2012arXiv1205.5510V}.  
Scaling off the Crab using Eq.~(\ref{eq:mdp}), this translates into 
a detection time of $\approx 10^{8}$ ksec. 
Therefore, this simple calculation shows that 
INTEGRAL is not well-suited 
to detect polarization from blazars. 

For other high-energy polarimeters, we determine 
the observation times using eqns.~(\ref{eq:sourcedomobstime}) and 
~(\ref{eq:bgddomobstime}). 
Tables ~\ref{table:PoGOLITE}~\ref{table:X-Calibur}~\ref{table:GEMS}~\ref{table:POLARIX}~\ref{table:ASTRO-H}~\ref{table:GAP} show approximate observation times 
required to detect polarization of 3C 279, PKS 1510-089 and 3C 454.3 
using the polarimeters, 
PoGOLITE, X-Calibur, GEMS-like missions, POLARIX, ASTRO-H and GAP. The scaling is done in terms of the photon 
number fluxes from observed energy fluxes. To do this computation, 
the effective areas for each polarimeter must be used to determine 
the counting rates, R for both Crab and the blazars. However, 
it is difficult to do this for future missions where the 
energy dependence of the effective areas is not understood very well yet. 
As described in the previous section, for our candidate blazars, 
this difference results in a factor of order unity, 
which is less than the factor between the background 
and signal dominated regimes.  The last 
column indicates the exposure times required to 
jointly measure the same polarization degree, $\Pi$ and position angle, $\psi$ 
 at $3\sigma$ above zero
in accordance with 
\citet{Weisskopf:2010se, 2013ApJ...773..103S}. 
The times  
increase in proportion to the square of 
the significance of the measurement. Thus, 
in going from a 3-sigma measurement to a 5-sigma measurement, 
the required time will increase by a factor of $\approx 2.8$ for 
a measurement of either 1 or 2 parameters. In going 
from a 3-sigma 1 parameter measurement to a 5-sigma 2 parameter measurement, 
this factor is $\pfrac{3^{2}}{4} \times \pfrac{5}{3}^{2} = 6.25$ 
All these 
times are subject to the assumption that 
systematic errors do not dominate the statistical errors. 

\subsubsection{Detection prospects and strategy}
From these estimates it is clear that 
the detection of polarization from blazars 
is challenging with balloon-borne polarimeters 
compared to space polarimeters. Space 
based instruments will almost certainly detect 
a signal, if indeed the high energy emission
 from jets in blazars is polarized. On the other hand, 
 balloon-borne experiments with a higher detection 
 time (equivalently a higher MDP threshold) will 
 need to have their systematics moderated 
 and perhaps need blazars with a higher flux and degree to 
 find a positive detection. 
Flaring blazars 
observed with dedicated balloon flights
may have a chance of having their polarization detected. 
This will demand planned observations based on triggers from 
the optical, UV, IR telescopes monitoring flaring activity. 
Our estimates 
 based on the MDP \citep{2013APh....41...63G, 2012arXiv1211.5094P} suggest X-Calibur 
 will have a better chance of detecting a polarization signal from 
 blazars than PoGOLITE.
 
With the future space based instruments, 
the sensitivity is much higher and 
thus statistically significant detections are highly likely 
with reasonable observing times. Space 
based instruments are not dedicated merely 
to polarization measurements and therefore, 
observational strategies are required. Due to the 
high sensitivity,  
quiescent or non-flaring blazars 
are also expected to be discovered. This 
is achievable during all sky scans as 
well as by targeted, longer observations of the more 
dim blazars. And for flaring blazars
optical, UV, IR triggers can be used for making 
detections. These strategies must be kept 
in mind while planning for future missions. 

\subsection{Motivation for synchronous multiwavelength polarimetry}
In addition to using published and archival data, it would 
be highly effective in to perform synchronous, multiwavelength polarimetry with 
high-energy polarimeters and low energy polarimeters suited for blazar
observations, such as  RoboPol \citep{robopol}, F-GAMMA
\citep{fgamma}, and OVRO \citep{ovro}. From the above results, idea of using a dedicated, 
specialised polarimeter 
at longer wavelengths in conjunction with 
a high energy polarimeter will significantly improve 
the chances of detecting high energy polarization from blazars. 
This is 
because monitoring the flaring activity of blazars as described in 
detail 
the next section
can 
be potentially be very useful to improve detection prospects. 
Furthermore, there is evidence that polarization
certainly at low energies is 
dynamic
\citep{2013arXiv1304.2819S}. This suggests 
that given low energy polarization 
is ultimately a source of high energy polarization, 
in the SSC scenario,  
polarimetric variability could be crucial at 
X-rays and gamma rays too. Polarimetric 
variability means that there are 
times when the degree of polarization 
increases and therefore 
detectability improves \citep{2011PASJ...63..489S}. 
In cases where the amount of polarization decreases 
with flux, variability in flux could yet be connected 
to that in polarization \citep{2011PASJ...63..639I, 2010Natur.463..919A}, a connection 
that would ultimately lead to clues about 
the underlying mechanisms. 
Thus,
multiwavelength polarimetry 
sharpen the emission models which connect the low 
energy to the high energy emission. Therefore, 
in order to faithfully probe the emission 
models, it is important to 
perform synchronised multiwavelength observations. 
Dedicated blazar polarimetric monitoring programs (such as RoboPol)
are thus critical to our understanding of polarization 
and in general emission processes in blazars. 

\section{Change in detectability due to flaring}
\label{sect:flaring}
\subsection{Different effects of flaring}
Blazars are well-known to flare at all
observed wavelengths, including in the gamma-ray \citep[e.g.,][and references therein]{2012arXiv1211.0274N}. 
The high energy (X-ray and gamma-ray)
fluxes can vary by up to a factor of 10 or 20 when they flare \citep[e.g.,][and references therein]{2012arXiv1211.0274N}. 
Of course, the factors could be higher for exceptional flares. 
Even the low energy fluxes have high flaring factors \citep{1997ARA&A..35..445U}.
As discussed in the previous section, 
flaring activity of blazars affect the detectability 
in the polarization domain. The direct reason 
is simply that those blazars in their 
quiescent state falling below the detection sensitivity 
of high energy polarimeters, 
may be pushed above threshold when they flare. 
In practice however, from the fluxes of our candidate blazars 
in the previous section and the observation times 
needed to detect them, it is only the very bright few 
blazars that will be detected anyway. And so the 
bright end of the source flux distribution $\dd{N}{F}$
is of interest. This is what we discuss in the next section~\ref{sect:fluxincrease}

Secondly, the polarization signal itself may vary 
due to the flaring activity. This could  
reduce the detectability if the degree of 
polarization goes down, as in the case of 
the gamma-ray flare seen from 3C 279 \citep{2010Natur.463..919A}.
Detectability also could be reduced to  due to 
variation in the polarization angle \citep{2008Natur.452..966M, 2010Natur.463..919A}. 
Alternately, the detectability could improve 
due to a positive correlation between 
flux and polarization, as reported in the case 
of AO 0235+164 and PKS 1510-089 by \citet{2011PASJ...63..489S}. 
This effect needs 
to be studied observationally. However, as 
shown in \citet{2011PASJ...63..639I}, there is 
no statistically significant correlation between 
the variation in optical and near-IR flux and variation in polarization 
properties 
namely polarization degree, $\Pi$ and 
position angle, $\psi$ of flaring blazars. Only 30$\%$ blazars 
in their well observed sample showed a positive correlation 
in the flux and polarization degree variations and 12$\%$ 
a negative correlation. And only 
3-4 blazars showed rotation of the polarization plane, 
with flaring. Also, there is 
no established statistical correlation between variations in 
gamma-ray flux and polarization properties. We will quantify the effect on an individual 
blazar that shows some correlation, later in this section.
However, this is not a significant effect for a statistical 
sample of blazars. Therefore our treatment will 
ignore this 
effect.

\subsection{Quantitative effect of flux increase on detectability}
\label{sect:fluxincrease}

Here we quantify the increase in the detected fraction of 
blazars purely due to an increase in their flux.
This effectively amounts to lowering 
of the detection limit by the same factor. For this  we use the 
formalism of Feldman and Pavlidou (in preparation) 
to compute this increased detection fraction 
detailed in Appendix~\ref{sect:effectofflaring}. 
This formalism models blazar high-energy output
with two states:
a quiescent state with flux $\fq$,
and a flaring state with flux $\ff = \etaf \fq$,
where the flaring enhancement factor is typically $\etaf \sim 10$. 
The duty cycle $\chi$ gives the fraction of time spent in the flaring state,
with typical values $\sim 20\%$. This means that the time-averaged mean 
flux $F = [1+(\etaf-1) \chi] \fq$ is
higher than the quiescent flux $\fq$.  Picking 
an average value of the flaring factor and duty cycle 
for the population of blazars
is an assumption which can easily be violated 
depending on the energy band of observation, the chosen blazar 
and indeed at different times for a single blazar. There is 
a large variance in flare properties of blazars and 
it is beyond the scope of this paper to delve into any 
of these complications. However, 
one can still make conservative predictions of 
increase in detectability of blazars by picking 
conservative values of flare factors and duty cycle. Higher 
flare factors and lower duty cycles will result in a 
positive increase in the detectability.

As mentioned before, polarization is likely to 
be detectable only
in the brightest blazars, and thus we study the effects of flaring
on the bright end 
of the source flux distribution. 
The LAT distribution \citep{2010ApJ...720..435A}
of blazar fluxes shows that the  
the number of sources above a mean flux $F$
is a broken power
law, with the two regimes divided at a
``break'' flux, $\fb$.
The brightest blazars
have $F > \fb$.
In this regime, the observed flux distribution is $dN/dF \propto F^{-\b1}$
where $\b1 = 2.49$, when the entire sample of blazars 
(i.e. both FSRQs and BL LACs) is taken.
Thus, the number of blazars detected by a high energy 
polarimeter in their quiescent state is
\beqar
\label{eq:sfdsimplified}
\ndet = \int_{\rm \fs}^{\infty} dF \dd{N}{F} \propto \pfrac{\fs}{\fb}^{1-\b1}
\eeqar
when the flux sensitivity 
is $\fs$ of the polarimeter. 

Now the effect of flaring is to 
effectively reduce the sensitivity limit by 
the flaring factor, $\etaf$from $\fs$ 
to $\fs / \etaf$. For a typical flaring factor $\etaf \sim 10$,
one would expect the number of sources to increase
by a factor of $\etaf^{\b1-1}$, or $\approx 1.5 $ orders of magnitude. 
On the other hand, the enhancement is only present for a fraction of
time given by the duty cycle.
This implies that effectively, one does not reduce the 
sensitivity by the full flaring factor $\etaf$, but instead reduces the 
sensitivity 
by a lower valued effective factor $\etaeff$
weighted by the duty cycle as given by 
Eq.~(\ref{eq:meantoflaring}). 
This reduces the total number of sources detected when flaring
to, 
\beqar
\label{eq:flaringeff}
\nfdet = \int_{\rm \fs/\etaeff}^{\infty} dF \dd{N}{F}  \propto \pfrac{\fs/\etaeff}{\fb}^{1-\b1}
\eeqar
For values, $\chi = 20\%$ and $\etaf = 10$, $\etaeff = 3.57$ 
which gives an increase factor of $ 6.67$ instead of 1.5 orders 
of magnitude. This implies that the 
number of additional sources detected above the 
sensitivity threshold due to flaring is $\approx$ 5-6 times 
the number if the sources were in their quiescent state. This is 
of course assuming that all sources can in principle flare 
and that the observation time is long enough all 
of these flares have been observed.  At any given instant, 
the number of additional sources is simply given by 
the product of the number of additional sources overall,
and the duty cycle $\chi$. Thus, the instantaneous boost for 
a duty cycle of $20\%$ is 1 additional source for every quiescent source. 

As there are a handful of blazars at the bright end of 
the source flux distribution to begin with, 
instead of a detailed statistical 
description, individual blazars 
that are known to flare and 
have a high degree of polarization at the lower energies 
should be selected for pointed observations. 
Individually, some blazars 
such as 3C 454.3 do show some correlation in the variation 
of their polarization properties with flaring. 3C 454.3 for instance
has exhibited a rotation in polarization angle with flaring. The 
reduction in the measured polarization degree could be as 
high as a factor of 100 from Table 2 in \citet{2011PASJ...63..639I} 
with a variation from $0.2 - 22.5 \%$. 
The variation in flux is also nearly a factor of 10 in V-band magnitudes. 
There is evidence for correlation over a day timescale 
between X-ray and gamma ray variability on 
one hand and optical variability on the other \citep{2010ApJ...715..362J}.
Thus, if the high energy fluxes also vary proportionally, the amplification 
in high energy flux is also a factor of 10. Scaling 
from Eq.~\ref{eq:mdp}, this could lead a factor of $\approx 25 - 250$ 
increase in the detection time, if there is perfect anti-correlation. 
On the other hand, given the nature of correlation found in 
\citet{2011PASJ...63..639I}, this is quite likely not such 
a drastic effect, and could even in some phases 
reduce the detection time. 

\section{Discussion and conclusions}
\label{sect:conc}

The prospects of detecting high energy 
polarization from blazars appear to be 
very realistic with current and 
future generations of polarimeters. The 
space based detectors such as POLARIX, ASTRO-H 
and a GEMS-like mission are quite likely to detect 
a polarization signal from blazars within 
reasonable observation times, based on their 
sensitivities, if the systematic uncertainties are not dominant. 
ASTRO-H for instance, 
could detect the listed blazars 
in $\sim 10^{6}$ seconds with a 
modest degree of polarization $O(10\%)$, 
and fluxes in a state of quiescence. A 
GEMS-like mission will take a few hours to do this. POLARIX  
has 
the potential to do the same in less than 
an hour. GAP is specifically designed 
for GRBs, however, it is quite possible that a flaring 
blazar whose polarization does not reduce 
with flux may be detected by it. Balloon borne polarimeters 
cannot do as well. Between the two listed here, X-Calibur is relatively 
the more sensitive, though our predictions suggest it will 
be challenging for it to provide a detection in reasonable exposure times. 
The key would be to observe a variable source with a positive 
correlation between flux and polarization degree in an episode of flaring. 

These observing prospects 
quantified in terms of the observation times required 
to make these detections are 
based on the MDP expression in Eq.~\ref{eq:mdp}. The 
underlying assumption is that the measurements  
are not limited by systematics and other observational constraints 
like mode of observation and high backgrounds near the 
source region. 
Therefore, the observation times 
may increase depending on these factors. However, 
our results provide an indication of the promise of 
blazar detections with several polarimeters particularly 
space based and the comparisons between them. 
Furthermore, we make the distinction between 
time required to measure the polarization degree alone 
from these blazars and that taken to 
produce joint measurements 
of the degree and the position angle \citet{Weisskopf:2010se, 2013ApJ...773..103S}. 
While a positive detection of a polarization signal 
from blazars 
will be a big achievement in adding a 
new source class to high energy polarimetry, 
a measurement of the degree will ultimately 
put constraints on the physics of jet emission. 

As described in section~(\ref{sect:flaring}), 
variability of blazar fluxes has a key role to 
play in the detection prospects. Statistically, 
the number of blazar detections are likely to increase with 
flaring events, possibly by a factor of 5-6 if we 
wait long enough allowing for all flaring candidates to flare. The instantaneous 
increase in sources on the other hand is $\approx 100\%$ 
for a duty cycle of $20\%$ or 1 additional blazar per quiescent 
blazar already detected. This is purely accounting for flux increase 
due to flaring. In practice, 
since the brightest blazars at the tail end of the flux distribution are 
the ones that are most likely to overcome the polarization detection 
threshold, this leads to a handful of additional blazars. Therefore, 
the detection strategy would be to plan observations 
of the brightest blazars with a history of flaring. A further 
selection can be made from these depending 
upon correlation between low energy polarization 
variations and high energy flux variability.  
Individual blazars, 
show an increase or reduction in 
detectability depending on variation of polarization properties with 
that in flux. 3C 454.3 for instance, may be more challenging to detect 
in polarization due to rotation of the polarization angle with flaring. 
Examples like this and their counterexamples strongly emphasise 
the need for synchronous, multiwavelength polarimetric observations. 
For this reason dedicated high-cadence optopolarimetric programs optimized 
for blazar detection are crucial to such observations. 
In general, the success of pointed blazar observations 
by high energy polarimeters could lie in the 
wealth of polarimetric 
and variability information provided 
observational studies at lower energies. 
Such observations will help select 
the most promising targets for blazar 
observations and also the 
times when they are most likely to be detectable 
based on variability. 

Not just episodic increases in flux, but 
persistent variability is also connected 
to polarimetry. A quantitative understanding of multiwavelength variability of 
blazars using statistical tools to determine timescales, 
correlation properties of variability constrains not only the 
size and location of the emission region, but also 
emission mechanisms, magnetic fields etc.  Thus, variability 
and polarimetric studies are complementary and 
should go hand in hand.

Polarimetry represents the next frontier in 
high energy astronomy with several scientific 
questions hinging on it. As indicated by 
several authors \citep[e.g.,.][]{Krawczynski:2011ym, 2013arXiv1303.7158K, 2013ApJ...774...18Z}, polarization 
can help to distinguish between different 
emission mechanisms of jets i.e. leptonic 
vs hadronic and also between different leptonic models. 
It is complementary to (non) detection of neutrinos from 
blazars. The magnetic field structure is directly probed 
by polarimetry \citep{Krawczynski:2011ym, 2013ApJ...774...18Z}. 
Measurements of 
high energy 
polarimetry are thus, a powerful probe of several 
astrophysical and physical questions. 

\acknowledgments{We would like to thank Felix Aharonian, Paolo Coppi, Markus Boettcher, Hirokazu Odaka for useful discussions on prospects of high energy polarimetry in general. We would also like to thank the anonymous referee for a detailed critical review that helped improve the paper. NC acknowledges support from his host institution, MPIK. VP acknowledges support by the ``RoboPol'' project, which is implemented under the ``Aristeia'' Action of the  ``Operational Programme Education and Lifelong Learning'' and is co-funded by the European Social Fund (ESF) and Greek National Resources, and by the European Comission Seventh Framework Programme (FP7) through grants PCIG10-GA-2011-304001 ``JetPop'' and PIRSES-GA-2012-31578 ``EuroCal''. The work of BDF was partially supported by NASA via the
Astrophysics Theory Program through award NNX10AC86G. 
}
\appendix
\section{Effect of flux increase due to flaring on source fluxes}
\label{sect:effectofflaring}
Every 
blazar flares given a long enough time. Feldman 
and Pavlidou treat the flaring and quiescent blazars 
as a single population with two states. Both 
flaring and quiescent blazars are represented by a single 
luminosity function. The two states differ by 
a flaring factor is given by 
\beqar
\label{eq:flaringfactor}
 \etaf= \frac{\ff}{\fq}
\eeqar
where $\ff$ and $\fq$ are fluxes in the flaring 
and quiescent state respectively. This flaring 
factor is assumed to be the same 
for all blazars and uniform in flux.  This is 
however an assumption that is very often untrue 
as there is substantial variance in the flare properties. 
Flare amplitudes may vary from a factor of a few to nearly 3-4 orders of 
magnitude in exceptional cases (Feldman and Pavlidou in preparation). 
This depends on the energy band of observation. Also, 
the same blazar can have a variation in the duty cycle 
and flaring amplitude \citep[e.g.,][]{2010ApJ...715..362J}. 
Our goal is not to delve into these details, but make some conservative estimates of the 
increase in detectability of blazars based on average properties.

Feldman and Pavlidou (in preparation) find that the source flux distribution 
fitting the \fermi-LAT data has contributions  
from both quiescent and flaring sources  
\citep{2010ApJ...720..435A}. This is at GeV energies. We need 
an extrapolation of the distribution 
down to the fluxes at keV and MeV energies 
at which the polarimeters operate in absence of observed 
source flux distribution at these energies. In practice, 
this is easy as blazars have a power-law spectrum. And 
this will be done at the end when we 
provide numbers for the fractional increase 
in the detected blazars. 

As a result of flaring, the break flux $\fb$ in the 
source flux distribution as observed in \fermi-LAT is shifted 
to the lower value $\fb / \etaf$. So, the source flux distribution 
is modified to include both flaring and quiescent blazars as 
 they may be treated as a single population given by, 
 \beqar
 \label{eq:sfdallblazars}
 \dndfo &=& A \pfrac{F}{\fb}^{-\b1} + A \etaf \cq \pfrac{\etaf F}{\fb}^{-\b1} ,  F \geq \fb 
\eeqar
since, we are only interested in the 
high end of the blazar flux distribution. 
Here $\cq$ represents the factor 
by which the number of blazars in the quiescent state 
exceed the number in the flaring state at any given time in 
the universe and is related to the fraction 
of the times the blazar is in the high state or 
the duty cycle, $\chi$ as $\cq = \frac{1-\chi}{\chi}$.
In this case,  the flaring flux $\ff$ is related to the mean flux, $\fm$ as, 
\beqar
\label{eq:meantoflaring}
 \ff &=& \frac{\etaf}{(\etaf-1) \chi + 1} \fm = \etaeff \fm
\eeqar
The mean flux is computed for the limit when all additional 
blazars that exceed 
the sensitivity limit due to flaring,  
are detected. For a duty cycle, $\chi = 20\%$ and 
a flaring factor, $\etaf = 10$, the effective flaring 
factor is $\etaeff = 3.57$.

Suppose, 
$\nun$ is the number of sources undetected 
in the quiescent state, but that are detected during 
their flaring state. Therefore, 
\beqar
\label{eq:nun}
\nun = \int_{\frac{\fs}{\etaeff}}^{\fs} \dd{N}{F} \ dF &=& \int_{\frac{\fs}{\etaeff}}^{\infty} \dd{N}{F} \ dF - \int_{\frac{\fs}{\etaf}}^{\infty} \dd{N}{F} \ dF \nn \\
&=& \nfdet - \nqdet 
\eeqar
where $\fs$ is the sensitivity of 
the polarimeter and $\etaeff$ is the 
effective amplication factor due to flaring 
accounting for a finite duty cycle. 
Therefore the fractional increase of sources observed 
due to flaring increases is given by, 
\beqar
\label{eq:ratioofnun}
f = \frac{\nun}{\nqdet}
\eeqar
where $\nqdet$ is the number of blazars detected 
in a state of quiescence i.e. without any increase in flux 
of any blazar due to flaring.

In accordance with 
\citep{2010ApJ...720..435A}. with  
the parameter values corresponding to the 
best fit values for all blazars both FSRQs and BL LACs put together 
are $A_{\rm \fermi} = 16.46\times10^{-14} {\rm cm^{2}\ sec\ deg^{-2}}$, 
$\fb = 6.60\times10^{-8} {\rm ph\ cm^{-2}\ sec^{-1}\ }$, $\b1 = 2.49$ 
and $\b2=1.58$. 
Picking a sensitivity 
higher than the break flux $\fs = \fb \etaeff$ 
as most certainly the blazars detectable are above the break, the 
value of $f \approx 5.7$ for  
a duty cycle, $\chi$ of $\sim20\%$ and a flaring factor of $\etaf = 10$ 
assuming all flaring events have occurred. Instantaneously, however, 
the gain is simply $\nun \times \chi$ which gives a 
factor of 2 for a uniform duty cycle of $20\%$. 
However, as stated 
before the actual numbers would depend 
on a number of factors including the assumption that 
all blazars flare. 

\bibliographystyle{apj}
\bibliography{blazarpol}
\end{document}